%% file: mixing_resub.tex
\documentclass[reprint,amsmath,amssymb,aps,onecolumn,superscriptaddress]{revtex4-2} 
\usepackage{bm}
\usepackage{hyperref}
\usepackage{color}
\pdfoutput=1

\input{definitions}

\begin{document}

\preprint{APS/123-QED}

\title{Mixing by squirmers in stratified fluids}
\author{Vaseem A. Shaik}
\email{vshaik@northwestern.edu}
\affiliation{Department of Mechanical Engineering, Institute of Applied Mathematics,\\
University of British Columbia, Vancouver, BC, V6T 1Z4, Canada}
\affiliation{Engineering Sciences and Applied Mathematics, Northwestern University, Evanston, IL 60208, USA}
\author{Gwynn J. Elfring}%
 \email{gelfring@mech.ubc.ca}
\affiliation{Department of Mechanical Engineering, Institute of Applied Mathematics,\\
University of British Columbia, Vancouver, BC, V6T 1Z4, Canada}

\date{\today}

\begin{abstract}
We find the mixing induced by a small swimming organism in a density stratified fluid. We model the swimmer as a spherical squirmer and quantify mixing through a mixing efficiency that is the ratio of rate of change of potential energy of fluid to the total work done by the swimmer. Assessing mixing in the near field and far field of the swimmer separately, we find that the near-field mixing aligns with past work, but the overall mixing is much larger than that caused by a point-sized swimmer (like force-dipole), although still small in weak stratification. Equivalent results are also obtained for a homogeneous dilute suspension of non-interacting swimmers. Our study highlights the impact of swimmer size on mixing, revealing that small-sized swimmers cannot induce substantial mixing. However, we propose few (unexplored) pathways through which small swimmers could induce significant mixing, ultimately contributing towards oceanic mixing.
\end{abstract}

\maketitle


\section{Introduction}
Small swimming organisms are ubiquitous in nature, inhabiting environments ranging from our gut and yogurt cultures to soil, lakes, ponds, and oceans. These organisms usually propel by converting internal energy or energy from surroundings to directed motion. They also display a plethora of non-trivial phenomena, including boundary accumulation \cite{Berke2008, Ezhilan2015}, upstream swimming in Poiseuille flows \cite{Hill2007, Kaya2009, Kaya2012, Kantsler2014, Peng2020}, motility-induced phase separation \cite{Cates2015}, and active turbulence \cite{Alert2022}. Significant progress has been made in understanding the motion of swimming organisms, driven by advancements in microfluidic technologies and the development of reduced-order models based on hydrodynamics and non-equilibrium statistical mechanics \cite{Brennen1977, Berg2004, Fauci2006, Lauga2009, Koch2011, Guasto2012, Elgeti2015, Lauga2016, Lauga2020}.

Swimming organisms, regardless of size, can mix their surrounding fluid. This so-called biogenic mixing has been thought to contribute towards the ocean mixing. Estimates of 1 TW mechanical energy contribution by marine life to ocean show that it is on par with 1 TW input from winds and $0.6-0.9$ TW from tides \cite{Munk1998,Dewar2006}. Also the $10^{-5} - 10^{-4} \, \rm{W /kg}$ energy production by most marine animals dominates over $10^{-9} - 10^{-8} \, \rm{W/kg}$ turbulent energy dissipation in the deep ocean \cite{Huntley2004, Kunze2006}, suggesting a possibility of significant biogenic mixing. But the turbulent mixing caused by the abundant $\mu$m-mm sized organisms has been shown to be nevertheless minimal \cite{Visser2007, gregg2009turbulence}. Mixing by such small organisms can, however, be induced by non-turbulent mechanisms, and be quantified by measuring: the fluid volume displaced by the organism \cite{Katija2009, Subramanian2010, Leshansky2010}; the diffusivity of a tracer in a suspension of swimming organisms \cite{Wu2000, Lin2011}; the vertical mass flux of salt or temperature responsible for density gradients in oceans \cite{Kunze2011}; or the fraction of the work done by organisms for swimming that is spent in mixing the fluid, the so-called mixing efficiency \cite{Katija2012}.

Swimming and the induced mixing in oceans, lakes, and ponds are inevitably affected by density gradients caused by temperature or salinity gradients. These density gradients are observed to hinder the dual vertical migration of dinoflagellates like \textit{C. Furca} and \textit{D. Acuta} \cite{Jephson2009}, and euphausiids like \textit{M. Norvegica} \cite{Bergstrom1997}.  Density gradients have also been identified as one of the contributing factors to the formation of algal blooms in Maude Weir pool \cite{Sherman1998}. They are also found to reduce the fluid volume dragged by swimming organisms, thereby reducing the mixing induced by them \cite{Shaik2020a,Shaik2020}.

Swimming organisms in vertical density gradients induce flows and propel at velocities different from those in homogeneous fluid. The flow due to a point-force or a force-dipole \cite{List1971, Ardekani2010}, as well as the far-field flow generated by a settling particle or a neutrally buoyant swimming organism in stratified fluids \cite{Shaik2020, Varanasi2022a}, is closed-streamlined and decays much faster than in homogeneous fluid. Swimmers in density gradients speed up or slow down relative to their speed in homogeneous fluids \cite{Doostmohammadi2012, Dandekar2019, More2020, Shaik2021} and also display a directed motion (or taxis termed densitaxis) by rotating to swim along or against or normal to the density gradients \cite{Shaik2024}.

The mixing efficiency of neutrally buoyant organisms swimming vertically in vertical density gradients has been well characterized for several minimal model organisms. For instance, the mixing efficiency of a swimming sheet that propels by passing traveling waves on its surface usually increases with stratification but decreases with fluid inertia, with a maximum efficiency for a sheet swimming in temperature-stratified water at negligible inertia \cite{Dandekar2019}. The mixing caused by a force-dipole representation of organisms that is accurate in the far field is negligible at negligible inertia assuming weak density advection and stratification. Also negligible is the ensemble averaged mixing due to dilute homogeneous suspension of force-dipoles with isotropic orientation distribution \cite{Wagner2014}. But the mixing induced by a spherical squirmer that propels due to a prescribed slip on its surface was found to be dominant in the near field where singularities other than force-dipole are equally important \cite{Shaik2021}. This mixing (assuming negligible inertia and weak advection) is asymptotically larger than that induced by the force-dipole, but still small for weak stratifications. Finite inertia and stratification tend to enhance the mixing induced by a single squirmer \cite{More2020} or even by a dilute suspension of squirmers \cite{Wang2015} but the maximum mixing induced is still small.

Here we find the mixing efficiency of a spherical squirmer swimming in an arbitrary direction in vertical density gradients. This way we unravel the effect of finite swimmer size on mixing, distinct from studies focusing primarily on point sized \cite{Wagner2014} or infinitely long swimmers \cite{Dandekar2019}, and also generalize the mixing induced by vertically swimming squirmers \cite{Shaik2021} to swimming in an arbitrary direction. We neglect fluid inertia and consider weak advection and stratification to find the mixing in both the near field and far field of the squirmer. The near-field mixing is in line with \cite{Shaik2021}, while the overall (near- and far-field) mixing is asymptotically larger than that induced by a force-dipole. We also discuss the use of the mixing efficiency of a single squirmer in finding the mixing by a dilute homogeneous suspension of squirmers, both without any orientational order or with order caused by densitaxis. We present the equations governing the flow and density fields surrounding the squirmer in Sec.~\ref{sec:model}, then analyze these fields in Sec.~\ref{sec:analysis} and the mixing efficiency in Sec.~\ref{sec:mixing}, and provide conclusions in Sec.~\ref{sec:conclusions}.

\section{\label{sec:model}Locomotion in density gradients}
We consider an organism swimming in an otherwise quiescent Newtonian fluid. The fluid density in the absence of organism, $\rho_{\infty}\lb( \bx \rb)$, exhibits spatial variations due to a similar variation in temperature or salinity. The density in oceans, lakes, and ponds exhibits a staircase pattern, characterized by relatively well-mixed layers separated by thinner regions with enhanced density gradients \cite{Tait1968, Neal1969, Howe1970, Tait1971, Neshyba1971, Phillips1972, Posmentier1977, Padman1987, Merryfield2000, Timmermans2008, Petropoulos2023}. But here for simplicity, we assume that the ambient (or background) density $\rho_{\infty}\lb( \bx \rb)$ varies linearly \cite{Ardekani2010, Doostmohammadi2012, Dandekar2019, More2020, Shaik2021, Shaik2024, Wagner2014, Wang2015}
\begin{equation}
    \frac{\bnabla \rho_{\infty} g}{\rhob} = N^2 \bgh,
\end{equation}
where the buoyancy or the Brunt-V\"{a}is\"{a}l\"{a} frequency $N$, is constant, is the pace at which a vertically displaced inviscid fluid element undergoes oscillations in density gradients, while $\bgh = \bg/\lb| \bg \rb|$ is a unit vector in the direction of gravity $\bg$ and $g  = \lb| \bg \rb|$ is the gravitational acceleration. Upon integration, we express the ambient density as $\rho_{\infty} = \rhob \lb( 1 + g^{-1} N^2 \bgh \cdot \lb( \bx - \bar{\bX} \rb) \rb)$, where $\rhob$ is the mean density of a region containing the organism that is significantly larger than the organism size, $\bar{\bX}$ is center of this region and $\bx$ denotes any position in space. These density variations induce ambient pressure variations $p_{\infty} \lb( \bx \rb)$ that, due to the quiescent nature of ambient fluid $\lb( \bu_{\infty} = \bzero \rb)$, satisfy the hydrostatic equation
\begin{equation}
    \bnabla p_{\infty} = \rho_{\infty} \bg.
\end{equation}

Introducing a swimming organism into the fluid typically disrupts the background density, as the organism’s properties, such as thermal conductivity or salt diffusivity, often differ from those of the fluid. Additionally, the organism induces a flow causing the transport of density. Hence the density field in the presence of an organism is written as
\begin{equation}
    \rho \lb( \bx \rb) = \rho_{\infty} \lb( \bx \rb) + \rho' \lb( \bx \rb),
\end{equation}
where $\rho'$ is the disturbance density. Similarly, we may express the pressure and fluid velocity in the presence of organism as $p = p_{\infty} + p'$, and $\bu = \bu'$, respectively.

Neglecting fluid inertia, the flow induced by organism satisfies the incompressible Stokes equations (in the Boussinesq approximation)
\begin{gather}
    \nabla \cdot \bu' = 0,\\
    -\bnabla p' + \rhob \nu \nabla^2 \bu' + \rho' \bg = \bzero,
\end{gather}
where $\nu$ is the kinematic viscosity of fluid. In the Boussinesq approximation \cite{Gray1976, Leal2007, Candelier2014}, density is treated as constant $\rhob$ in all the terms in equations governing flow, except for the body force. Also the fluid viscosity and temperature or salt diffusivity are considered to be constant.

The transport of temperature or salinity is governed by an advection-diffusion equation, which in turn determines the density transport. Within the Boussinesq approximation, density changes are assumed to be linearly proportional to changes in temperature or salinity, hence the density transport is also governed by an advection-diffusion equation with diffusivity $\kappa$
\begin{equation}
    \frac{\partial \rho'}{\partial t} + \bu' \cdot \nabla \rho' + \frac{\rhob}{g} N^2 \bu' \cdot \bgh = \kappa \nabla^2 \rho'.
    \label{eqn:adv-diff}
\end{equation}
This equation reveals advection of disturbed density $\bu' \cdot \nabla \rho'$ and background density $ \frac{\rhob}{g} N^2 \bu' \cdot \bgh$ by the disturbance flow. 

Far from the swimming organism, the disturbance caused by it diminshes, hence
\begin{equation}
    \bu' \to \bzero, \rho' \to 0 \,\, {\rm{as}} \,\, r = \left| \br \right| \to \infty,
\end{equation}
where $\br = \bx - \bX$ and $\bX$ denotes the position of center of organism. While an organism's surface is usually impermeable to salt or insulating against heat transfer, hence due to the linearity between these scalar fields and the density, we expect the density near organism satisfies a no-flux boundary condition, $\bn \cdot \nabla \rho = 0 \,\, {\rm{for}} \,\, \bx\in \partial\fB$, which in terms of disturbance variables simplifies to
\begin{equation}
    \bn \cdot \nabla \rho' = -  \frac{\rhob}{g} N^2 \bn \cdot \bgh \,\, {\rm{for}} \,\, \bx\in \partial\fB.
\end{equation}
Here $\partial\fB$ denotes the surface of the organism, and $\bn$ an outward pointing unit normal to it. Also the fluid velocity near the organism adheres to the usual no-slip boundary condition. As the organism propels with some unknown translational velocity $\bU$ and angular velocity $\bOmega$ due to a velocity $\bu^s$ on its surface caused by activity, the fluid velocity satisfies
\begin{equation}
    \bu' = \bU + \bOmega \times \br + \bu^s \,\, {\rm{for}} \,\, \bx\in \partial\fB.
    \label{eqn:no-slip}
\end{equation}

We model the swimming organism as a spherical squirmer of radius $a$. This model effectively captures the locomotion of ciliated organisms like \textit{Paramecium} and \textit{Opalina}, which move through coordinated ciliary beating on their surface, as well as the motion of phoretic particles propelled by surface chemical reactions. In this model, we maintain a fixed swimmer shape and express its activity through a prescribed effective slip on the swimmer's surface \cite{Lighthill1952, Blake1971, Ishikawa2006}
\begin{equation}
    \bu^s = -\sum_n \frac{2}{n\lb(n+1\rb)} B_n P_n'\lb( \bp \cdot \bn \rb) \bp \cdot \lb( \bI - \bn \bn \rb).
\end{equation}
Here $\bp$ is the swimmer orientation, $B_n$s are called the squirming modes, $P_n$ is the Legendre polynomial of degree $n$ and $P_n' \lb( x \rb) = \frac{d}{dx} P_n \lb( x \rb)$. In homogeneous Newtonian fluids, the $B_1$ mode exclusively determines the swim speed, $U_N = \frac{2}{3} B_1$, while the $B_2$ mode governs the slowest decaying flow and, consequently, establishes the far-field representation of the organism. We denote the ratio of these two modes by $\beta = B_2/ B_1$, where $\beta > 0$ for pullers (like \textit{C. reinhardtii}) that swim by pulling the fluid in front of them but $\beta < 0$ for pushers (like \textit{E. coli}) which propel by pushing the fluid behind them. Here we retain only these first two squirming modes in line with the common practice with studying swimming under confinement \cite{Li2014, Shaik2017}, in complex fluids \cite{Zhu2011, Zhu2012, Li2014a, Yazdi2014, Yazdi2015, Yazdi2017}, and at finite inertia \cite{Wang2012, Khair2014, Chisholm2016}.

Neglecting organism’s inertia, the net force and torque acting on it should vanish
\begin{gather}
    \bF = \int_{ \partial\fB} \bn \cdot \bsigma' dS + \int_{\fV_p} \lb( \rho_p - \rho_{\infty} \rb) \bg \, dV = \bzero, \label{eqn:force-free}\\
    \bL = \int_{ \partial\fB} \br \times \lb( \bn \cdot \bsigma' \rb) dS + \int_{\fV_p} \br \times \lb( \rho_p - \rho_{\infty} \rb) \bg \, dV = \bzero, \label{eqn:torque-free}
\end{gather}
where $\bsigma' = -p' \bI + \rhob \nu \lb( \nabla \bu' + \lb( \nabla \bu' \rb)^T \rb)$ is the disturbance stress, while $\fV_p$ and $\rho_p$ denote the volume and density of the organism, respectively. The force and torque have two constituents – a hydrodynamic one due to the flow induced by the organism and a hydrostatic one due to the mismatch in density between the organism and the background fluid, and any asymmetries in the shape or mass distribution of the organism (e.g., bottom-heaviness). Generally, aquatic organisms have density similar to their suspending fluid $\rho_p \approx \rhob$. But for sake of simplicity, we posit here that the organism's density is always identical to the fluid density $\rho_p = \rho_{\infty}$ \cite{Shaik2021,Lee2019,Varanasi2022}. Consequently, as the organism swims and explores different density environments, it continuously adjusts its density to remain density matched with the background fluid. This regulation of density can be achieved through various means, such as utilizing a swim bladder \cite{Villareal2003} or employing mechanisms like carbohydrate ballasting \cite{Boyd2002} or ion replacement \cite{Sartoris2010}.

We non-dimensionalize the variables with relevant scales: the characteristic organism size $a$, the density gradients across the organism $\rhob N^2 a /g$, the swimming speed in homogeneous fluid $U_N$, and the viscous pressure or stress scale $\rhob \nu U_N/a$. Using the same symbols for dimensionless variables as their dimensional counterparts, we arrive at the dimensionless governing equations
\begin{gather}
    \nabla \cdot \bu' = 0, \label{eqn:continuity-ND}\\
    -\bnabla p' + \nabla^2 \bu' + Ri_v \rho' \bgh = \bzero, \label{eqn:Stokes-ND}\\
    Pe \lb( \frac{\partial \rho'}{\partial t} + \bu' \cdot \nabla \rho' + \bu' \cdot \bgh \rb) = \nabla^2 \rho'. \label{eqn:density-ND}
\end{gather}
These equations and the underlying physics are dictated by two dimensionless numbers: the viscous Richardson number $Ri_v = a^3 N^2 / \nu U_N$ that measures the ratio of buoyancy to viscous forces, and the P\'eclet number $Pe = a U_N / \kappa$ which quantifies the ratio of density advection to diffusion. The viscous Richardson number can be expressed as $Ri_v = Re/Fr^2$, with the Reynolds number $Re = a U_N / \nu$ being the ratio of inertia to the viscous forces and the Froude number of swimmer $Fr = U_N/N a$ the ratio of inertia to buoyancy forces. Also the P\'eclet number can be written as $Pe = Re Pr$, with the Prandtl number $Pr = \nu / \kappa$ being the ratio of momentum to mass diffusivity. 

We determine typical values of these dimensionless numbers by examining the characteristics of planktonic organisms and oceanic water. Plankton like \textit{E. coli}, \textit{Paramecium}, and \textit{Daphnia} have sizes $a$ ranging from 1 $\mu$m to 5 mm and swim in water at speeds  $U_N$ varying from 10 $\mu$m$/\rm{s}$ to 1 cm$/\rm{s}$. In contrast, water viscosity $\nu \approx 10^{-6}$  m$^2/$s, and its Prandtl number $Pr$ is either 7 or 700 for temperature- and salinity-induced density variations, respectively. Additionally, buoyancy frequency $N$ in various aquatic environments ranges from $10^{-4}$  to 0.3 s$^{-1}$. Using these parameters, we estimate that the Reynolds number $Re$ lies in the range $10^{-5}-50$, the viscous Richardson number $Ri_v$ varies from $10^{-15}-1$, while the P\'eclet number $Pe$ ranges from $7 \times 10^{-5} - 350$ or $7 \times 10^{-3}-3.5 \times 10^{4}$, depending on whether the density variations are temperature- or salinity-induced.

Here we work in an inertialess realm, the limit $Re \to 0$, relevant to the motion of small $\lb(\sim 10-1000 \,\, \mu\rm{m \,\, sized} \rb)$ organisms. We also assume the disturbance density field is steady when viewed from a reference frame translating with the organism, represented as 
\begin{equation}
    \frac{D \rho'}{Dt} = \lb( \bu' - \bU \rb) \cdot \nabla \rho'.
\end{equation}
Because of this assumption, the density transport equation \eqref{eqn:density-ND} simplifies to
\begin{equation}
    \nabla^2 \rho' = Pe \lb( \lb( \bu' - \bU \rb) \cdot \nabla \rho' + \bu' \cdot \bgh \rb).
    \label{eqn:density-NDFinal}
\end{equation}
We find the flow and density disturbances induced by the spherical squirmer at weak density advection and stratification.

\section{\label{sec:analysis}Analysis}
At weak advection, $Pe \ll 1$, density transport is governed by diffusion at leading order $\nabla^2 \rho' = 0$ along with the no-flux boundary condition on the surface of organism. The density field that satisfies these constraints and decays far from the organism is \cite{King2003}
\begin{equation}
    \rho'_{\rm{diff}} = \frac{\br \cdot \bgh}{2r^3}.
    \label{eqn:diff-density}
\end{equation}
This diffusive solution implies that the disturbance density is not homogeneous on the surface of organism $\lb(r=1\rb)$, as the vector in the radial direction ${\br}$ varies over the surface. Also the density $\rho = \rho_{\infty} + \rho'$ is not homogeneous on the surface of organism. But the density distribution \eqref{eqn:diff-density} breaks down far from the organism as no matter how small $Pe$, advection becomes as important as diffusion at the screening length $r = l_{\rho} \sim \frac{1}{Pe} \gg 1$. Beyond this distance, both advection and diffusion need to be accounted for. 

Similarly at weak stratification, $Ri_v \ll 1$, buoyancy forces can be neglected relative to viscous forces in the Stokes equations. Hence the velocity field in stratified fluids is the same as that in homogeneous fluids to leading order in $Ri_v$. This velocity field is well known \cite{Blake1971, Ishikawa2006} and it decays as $1/r^2$. But this velocity description breaks down far from the organism as no matter how small $Ri_v$, buoyancy forces become the same order of magnitude as the viscous forces at the stratification screening length $r = l_s$, where $l_s \sim \lb( Ri_v Pe \rb)^{-1/4}$ for $Pe \ll Ri_v^{1/3}$ but $Ri_v^{-1/3}$ for $Pe \gg Ri_v^{1/3}$ \cite{Shaik2024}. Beyond $l_s$, both buoyancy and viscous forces need to be considered.

Determining the velocity and density fields far from the organism involves solving \eqref{eqn:continuity-ND}, \eqref{eqn:Stokes-ND}, and \eqref{eqn:density-NDFinal}, considering both advection and diffusion in \eqref{eqn:density-NDFinal}, and viscous and buoyancy forces in \eqref{eqn:Stokes-ND}, along with some additional forcing terms to account for the presence of a force-free and torque-free swimming organism in the far field. Specifically, a Stresslet $\bS$ for the disturbance flow induced by the swimming organism, and a concentration-dipole $\bD$ for the disturbance density
\begin{gather}
    \nabla \cdot \bu' = 0, \label{eqn:FF-continuity}\\
    -\nabla p' + \nabla^2 \bu' + Ri_v \rho' \bgh + \bS \cdot \nabla \delta \lb( \br \rb) = \bzero,\\
    -Pe \bU_N \cdot \nabla \rho' + Pe \bu' \cdot \bgh = \nabla^2 \rho' + \bD \cdot \nabla \delta \lb( \br \rb). \label{eqn:FF-density}
\end{gather}
For deriving \eqref{eqn:FF-density} from \eqref{eqn:density-NDFinal}, we not only added the concentration-dipole term, but also linearized \eqref{eqn:density-NDFinal} by neglecting the nonlinear term $\bu' \cdot \nabla \rho'$ and approximating the swimmer velocity with its value in homogeneous fluids, $\bU \equiv \bU_N = \bp$. Neglecting the nonlinear term $\bu' \cdot \nabla \rho'$ compared to the linear one $\bu' \cdot \bgh$ is justified in the far-field of the swimmer. Because as all the disturbance variables decay with the distance from the swimmer, disturbance variables are small in the far-field, and hence their product can be neglected. On the other hand, approximating swimmer velocity with its value in homogeneous fluids is justified for weak density gradients, a limit relevant to our work. Specifically, the swimmer velocity in weak density gradients can be written as a small correction to its value in homogeneous fluids, with the scaling of the correction with $Ri_v$ and $Pe$ depending on the relative order of magnitude of screening lengths $l_s /l_{\rho}$ \cite{Shaik2024}. In a similar manner, $\bS$ and $\bD$ can be approximated as being independent of $Ri_v$ and $Pe$ for weak density gradients and density advection,
hence $\bS = 2\pi \beta \lb( 3 \bp \bp - \bI \rb)$ \cite{Ishikawa2006} and $\bD = -2 \pi \bgh$ \cite{Varanasi2022}. 

\section{\label{sec:mixing}Mixing efficiency}
Any flow induced displacement of the density field changes the potential energy of the fluid. But a distinction must be made between the reversible (referred to as stirring) and irreversible (referred to as mixing) changes in the potential energy \cite{Villermaux2019, Peltier2003, Caulfield2021}. These changes are reversible in the absence of diffusion, as the perturbed density field may return to its equilibrium position by reversing the flow or due to gravity.  But the potential energy changes are irreversible and the potential energy is non-decreasing in time, if account is made of the diffusion that acts on small length scales to smooth out the density gradients.

The mixing efficiency $\eta$ has been usually defined as the ratio of the irreversible increase in (the appropriately defined background) potential energy of the fluid due to mixing, and sum of this increase and the irreversible conversion of kinetic energy to internal energy through viscous dissipation, which is again non-decreasing in time \cite{Peltier2003, Caulfield2021}. This standard definition implies $\eta \in [0, 1]$. But here we do not differentiate between mixing and stirring and say that mixing efficiency is the ratio of the changes in potential energy of the fluid (could be increasing or decreasing in time) and the sum of these changes in potential energy and the viscous dissipation. This is why the mixing efficiency can be negative in our case. Although our definition of mixing efficiency differs from the aforesaid usual definition used in oceanic or stratified mixing literature \cite{Peltier2003, Caulfield2021}, it aligns with the literature on swimmer induced mixing \cite{Wagner2014, Wang2015, Dandekar2019, More2020}, and mixing estimates different from these two have been also used in the literature. See Table 1 in Ref. \cite{Gregg2018} for a list of various quantities used to estimate oceanic mixing.

We apply these concepts to calculate mixing by swimming organisms. Swimmers have to do work on the fluid to overcome viscous resistance and propel. In the absence of density gradients, work done by the organism is equal to viscous dissipation in the fluid,
\begin{equation}
    - \int_{\partial \fB} \bn \cdot \bsigma' \cdot \bu' \, dS = \frac{1}{2} \int_{\fV} \bgammad' : \bgammad' \, dV,
\end{equation}
where $\bgammad' = \nabla \bu' + \lb( \nabla \bu' \rb)^T $ and $\fV$ denotes the fluid volume outside the organism. In density gradients, however, some of the work is spent in displacing constant density surfaces (or isopycnals) from the horizontal and the ensuing changes in the potential energy of the fluid $\lb(\Delta PE \rb)$, 
\begin{equation}
    - \int_{\partial \fB} \bn \cdot \bsigma' \cdot \bu' \, dS = - Ri_v \int_{\fV}{\rho' \bgh \cdot \bu' \, dV} + \frac{1}{2} \int_{\fV} \bgammad' : \bgammad' \, dV.
\end{equation}
The mixing efficiency $\eta$ measures the fraction of work done by the organism that is spent in changing the potential energy of the fluid 
\begin{equation}
    \eta = \frac{- Ri_v \int_{\fV}{\rho' \bgh \cdot \bu' \, dV}}{- \int_{\partial \fB} \bn \cdot \bsigma' \cdot \bu' \, dS}.
    \label{eqn:mixing-def}
\end{equation}
Here we evaluate the mixing efficiency at weak advection and stratification.

Enforcing the no-slip condition \eqref{eqn:no-slip} and incorporating the force-free and torque-free constraints \eqref{eqn:force-free}, \eqref{eqn:torque-free}, the work done in the denominator can be written as $- \int_{\partial \fB} \bn \cdot \bsigma' \cdot \bu' \, dS = - \int_{\partial \fB} \bn \cdot \bsigma' \cdot \bu^s \, dS$. We see that this work done depends on the stress field near the swimmer, which to leading order in $Ri_v$, is the same as that in the homogeneous fluids. Hence evaluating the work done with the stress field in homogeneous fluids yields $- \int_{\partial \fB} \bn \cdot \bsigma' \cdot \bu^s \, dS = 6 \pi \lb( \beta^2 + 2 \rb)$, ultimately simplifying the mixing efficiency to
\begin{equation}
    \eta = \frac{- Ri_v \int_{\fV}{\rho' \bgh \cdot \bu' \, dV}}{6 \pi \lb( \beta^2 + 2 \rb)}.
    \label{eqn:eta-final}
\end{equation}

We also evaluate the potential energy changes using the leading order diffusive density, $\rho'_{\rm{diff}}$ from \eqref{eqn:diff-density}, and the velocity field in homogeneous fluids \cite{Blake1971, Ishikawa2006}
\begin{equation}
    {\bu}' = - \frac{{\bp}}{2r^3}  + \frac{3}{2 r^3} \frac{{\bp} \cdot {\br \br}}{r^2} + \frac{3 \beta}{2} \left( \frac{1}{r^4} - \frac{1}{r^2} \right) \left( - \frac{1}{2} + \frac{3}{2} \left( \frac{{\bp} \cdot {\br}}{r} \right)^2 \right) \frac{{\br}}{r} + \frac{3 \beta}{2r^4} \left( \frac{ {\bp} \cdot {\br} }{r} \right) \left( \frac{ {\bp} \cdot {\br} }{r} \frac{{\br}}{r} - {\bp}  \right).
\end{equation}
This calculation reveals that the integral in $\Delta PE$ converges with $r$, hence the integral is dominated by the region near the swimmer $r \sim O\lb( 1 \rb)$, giving the near field $\Delta PE$ as
\begin{equation}
     \Delta PE_{\rm{near}} = \frac{\pi Ri_v \beta \lb( 3 \lb( \bp \cdot \bgh \rb)^2 - 1 \rb)}{5}.
\end{equation}
Recall the flow in homogeneous fluids is a superposition of different singularities, although all singularities are important in the near field, due to symmetry, only the stresslet flow $\bu_{SS}$ yields finite potential energy changes in the near field, where
\begin{equation}
    {\bu}_{SS} = - \frac{3 \beta}{2r^2} \left( - \frac{1}{2} + \frac{3}{2} \left( \frac{{\bp} \cdot {\br}}{r} \right)^2 \right) \frac{{\br}}{r}.
\end{equation}
Said differently, the near-field $\Delta PE$ is caused by the displacement of diffusive density by the stresslet flow induced in homogeneous fluids, which also explains $\Delta PE$ dependence on the swimming orientation $\bp$ and the linear dependence on the squirming ratio $\beta$.

Enforcing the near-field $\Delta PE$ in \eqref{eqn:eta-final}, we derive the near-field mixing efficiency
\begin{equation}
    \eta_{\rm{near}} = \frac{Ri_v \beta}{30 \lb( \beta^2 + 2\rb)} \lb( 3 \lb( \bp \cdot \bgh \rb)^2 - 1 \rb),
    \label{eqn:mixing-near}
\end{equation}
which aligns with Ref.~\cite{Shaik2021} for vertical swimmer orientations. The mixing efficiency generally depends on the swimmer orientation and if the swimmer is a pusher or a puller. Vertically swimming organisms $\lb( \bp \parallel \bgh \rb)$, for instance, mix the fluid twice as strongly as the horizontal swimmers $\lb(\bp \perp \bgh \rb)$ simply because the stresslet flow in vertical direction $\lb( \bu_{SS} \cdot \bgh \rb)$ induced by vertical swimmers is twice as much as that by the horizontal ones. Also pullers mix the fluid in an opposite sense to pushers. Pullers through their stresslet contribution pull the fluid towards themselves along their axis and eject fluid sideways, while pushers do the opposite. So if the pullers displace heavier fluid (relative to the background density) upwards or lighter fluid downwards, ultimately rising the potential energy of the fluid, pushers displace fluid oppositely and lower the potential energy.


There could also be an additional contribution to mixing efficiency arising from the far field of the swimming organism $\lb( r \gg 1 \rb)$. We find this far-field mixing efficiency by evaluating the potential energy changes with the far-field density and velocity fields satisfying \eqref{eqn:FF-continuity}-\eqref{eqn:FF-density}. These far-field variables and the associated mixing, unlike the near-field mixing, are affected by the density advection or stratification, hence they depend on the relative order of magnitude of screening lengths $l_s$ and $l_{\rho}$.

Generally, we can divide the region surrounding the swimmer into three layers depending on where the advection or buoyancy matter \cite{Shaik2024}. But here for convenience, we define $r<l_s$ as the inner region and carefully account for any density advection effects in this region, while $r>l_s$ denotes the outer region. The leading order diffusive density and velocity fields that characterized the near-field mixing are nothing but the leading order variables in the inner region. Then finding the leading order density and velocity fields in the outer region, we can determine the far-field mixing efficiency. Although we can solve \eqref{eqn:FF-continuity}-\eqref{eqn:FF-density} for any value of $l_s/l_{\rho}$, analytical evaluation of mixing efficiency is possible in the cases where these screening lengths are well separated $l_s \ll l_{\rho}$, and $l_s \gg l_{\rho}$.

\subsection{$l_s \ll l_{\rho}$}
In this limit, density transport is governed by diffusion in both inner and outer regions for $r < l_{\rho}$. Then the functional form of density in the matching region, $ r \sim l_s \gg 1$, can be found by balancing the diffusion with the advection of background density by the disturbance velocity
\begin{equation}
    \nabla^2 \rho' \sim Pe \bu' \cdot \bgh.
    \label{eqn:density-match-lim1}
\end{equation}
The advection of disturbed density with particle velocity $Pe \, \bU_N \cdot \nabla \rho'$ becomes important only for $ r > l_{\rho} \gg l_s$ and hence is negligible in the matching region $r \sim l_s$. Also the disturbance velocity far from a force-free and torque-free swimmer is that of a stresslet, $\bu' = \bu_{SS} \sim 1/r^2$ for $r \sim l_s \gg 1$. Using $\bu' \sim 1/r^2$ in \eqref{eqn:density-match-lim1}, we derive that $\rho' \sim Pe $. This scaling exceeds the diffusive density for $r > Pe^{-1/2}$ and can be used to find the stratification screening length by balancing buoyancy forces with viscous forces near $r \sim l_s$
\begin{equation}
    \nabla^2 \bu' \sim Ri_v \rho' \bgh.
\end{equation}
We find that $l_s \sim \lb( Ri_v Pe \rb)^{-1/4}$ and hence the condition $l_s \ll l_{\rho}$ implies that $Pe \ll Ri_v^{1/3}$.

We find density and velocity fields in the outer region to ultimately determine the far-field mixing. We first rescale these variables by requiring that the rescaled variables are $O\lb( 1 \rb)$ in the matching region $\lb( r \sim l_s \rb)$. As $\rho' \sim Pe$, $\bu' \sim 1/r^2$, and $p' \sim 1/r^3$ near $r \sim l_s \gg 1$, we rescale them, denoting the rescaled variables with tilde, as $r = \rt/\lb( Ri_v Pe \rb)^{1/4}$, $\rho' = Pe \rhot'$, $\bu' = \lb( Ri_v Pe \rb)^{1/2} \but'$, and $p' = \lb( Ri_v Pe \rb)^{3/4} \pt'$. Hence, the far-field equations \eqref{eqn:FF-continuity}-\eqref{eqn:FF-density} written in terms of rescaled variables are
\begin{gather}
    \nablat \cdot \but' = 0, \label{eqn:FF-continuity-lim1}\\
    - \nablat \pt' + \nablat^2 \but' + \rhot' \bgh + \bS \cdot \nablat \delta \lb( \brt \rb) = \bzero,\\
    \nablat^2 \rhot' + \frac{Pe^{3/4}}{Ri_v^{1/4}} \bU_N \cdot \nablat \rhot' - \but' \cdot \bgh = \sqrt{\frac{Ri_v}{Pe}} \bD \cdot \nablat \delta \lb( \brt \rb). \label{eqn:FF-density-lim1}
\end{gather}
Similarly the far-field mixing efficiency in terms of rescaled variables is
\begin{equation}
    \eta_{\rm{far}} = -\frac{\lb( Ri_v Pe \rb)^{3/4}}{ 6 \pi \lb( \beta^2 + 2 \rb) } \int_{\tilde{\fV}} {\rhot' \bgh \cdot \but' \tilde{dV}}.
\end{equation}
To prevent double counting the near-field mixing efficiency, $- \frac{Ri_v}{6\pi \lb( \beta^2 + 2 \rb)} \int_{\fV}{\rho'_{\rm{diff}} \, \bgh \cdot \bu_{SS} \, dV}$, we rescale it and subtract it from the far-field mixing efficiency deriving
\begin{equation}
    \eta_{\rm{far}} = -\frac{\lb( Ri_v Pe \rb)^{3/4}}{ 6 \pi \lb( \beta^2 + 2 \rb) } \int_{\tilde{\fV}} {\lb( \rhot' \bgh \cdot \but' - \rhot'_{\rm{diff}} \bgh \cdot \but_{SS} \rb) \tilde{dV}},
\end{equation}
where $\rhot'_{\rm{diff}} = \sqrt{\frac{Ri_v}{Pe}} \frac{\brt \cdot \bgh}{2 \rt^3}$.

Neglecting the advection of disturbed density by particle velocity, $\frac{Pe^{3/4}}{Ri_v^{1/4}} \bU_N \cdot \nablat \rhot'$ for $Pe \ll Ri_v^{1/3}$, we see from \eqref{eqn:FF-continuity-lim1}-\eqref{eqn:FF-density-lim1} that $\rhot'$ and $\but'$ are linear in $\bS$ and $\sqrt{\frac{Ri_v}{Pe}} \bD$ or in $\beta$ and $\sqrt{\frac{Ri_v}{Pe}}$.  Consequently as $\eta_{\rm{far}} / \lb( Ri_v Pe \rb)^{3/4}$ depends on the product of $\rhot'$ and $\but'$, it should be linear in $\beta^2$, $\beta \sqrt{\frac{Ri_v}{Pe}}$, and $\frac{Ri_v}{Pe}$ or $\eta_{\rm{far}}$ should be linear in $\beta^2 \lb( Ri_v Pe \rb)^{3/4}$, $\beta Ri_v^{5/4} Pe^{1/4}$, and $Ri_v^{7/4} Pe^{-1/4}$. The $O\lb( Ri_v Pe \rb)^{3/4}$ far-field mixing was established previously by modeling the swimming organisms as force-dipoles \cite{Wagner2014} or as spherical squirmers with a negligible concentration-dipole relative to the stresslet (valid for $Pe \gg Ri_v$) \cite{Shaik2021}. But this far-field mixing is negligible compared to the $O\lb(Ri_v \rb)$ near-field mixing. Only the $O\lb( Ri_v^{7/4} Pe^{-1/4} \rb)$ far-field mixing could be of the same order of magnitude as the near-field mixing.

We neglect $\frac{Pe^{3/4}}{Ri_v^{1/4}} \bU_N \cdot \nablat \rhot'$ and solve \eqref{eqn:FF-continuity-lim1}-\eqref{eqn:FF-density-lim1} in Fourier space to derive the Fourier transform of density and velocity fields
\begin{gather}
    \rhoh'\lb(\bk\rb) = - \frac{6\pi \beta i \lb( \bp \cdot \bk \rb) \lb( k^2 \lb( \bp \cdot \bgh \rb) - \lb( \bp \cdot \bk \rb) \lb( \bk \cdot \bgh \rb) \rb) }{ k^6 + k^2 - \lb( \bk \cdot \bgh \rb)^2 } - \sqrt{\frac{Ri_v}{Pe}} \frac{ 2 \pi i k^4 \lb( \bk \cdot \bgh \rb) }{  k^6 + k^2 - \lb( \bk \cdot \bgh \rb)^2  }, \label{eqn:density-Fourier-lim1}\\
    \bgh \cdot \buh' \lb( \bk \rb) = \frac{ 6 \pi \beta i k^2 \lb( \bp \cdot \bk \rb) \lb( k^2 \lb( \bp \cdot \bgh \rb) - \lb( \bp \cdot \bk \rb) \lb( \bk \cdot \bgh \rb) \rb) }{  k^6 + k^2 - \lb( \bk \cdot \bgh \rb)^2 } - \sqrt{\frac{Ri_v}{Pe}} \frac{ 2 \pi i \lb( \bk \cdot \bgh \rb) \lb( k^2 - \lb( \bk \cdot \bgh \rb)^2 \rb) }{ k^6 + k^2 - \lb( \bk \cdot \bgh \rb)^2 }.
    \label{eqn:velocity-Fourier-lim1}
\end{gather}
Here we denote the Fourier transform of any variable $\ft \lb( \brt \rb)$ by $\fh \lb( \bk \rb)$, and define the Fourier and inverse Fourier transforms as
\begin{equation}
    \fh \lb( \bk \rb) = \int{ \ft \lb( \brt \rb) e^{- i \bk \cdot \brt} d\brt }, \thickspace \thickspace \ft \lb( \brt \rb) = \frac{1}{8\pi^3} \int{ \fh \lb( \bk \rb) e^{ i \bk \cdot \brt } d\bk }.
\end{equation}
We also transform the integral in $\eta_{\rm{far}}$ from real space to Fourier space by extending the integration volume to include particle volume $\lb( \fV_p \rb)$, and then using the convolution theorem to derive
\begin{equation}
    \eta_{\rm{far}} = - \frac{ \lb( Ri_v Pe \rb)^{3/4} }{ 48 \pi^4 \lb( \beta^2 + 2 \rb) } \int{ \lb( \rhoh'\lb( \bk \rb) \bgh \cdot \buh'\lb( -\bk \rb) - \rhoh'_{\rm{diff}} \lb( \bk \rb) \bgh \cdot \buh_{SS} \lb( -\bk \rb) \rb) d\bk }.
\end{equation}
Here $\rhoh'_{\rm{diff}} \lb( \bk \rb) = - \sqrt{\frac{Ri_v}{Pe}} \frac{2\pi i \lb( \bk \cdot \bgh \rb) }{ k^2 }$ and $\bgh \cdot \buh_{SS} \lb( \bk \rb) = \frac{6\pi i \beta \lb( \bp \cdot \bk \rb)}{k^4} \lb( k^2 \lb( \bp \cdot \bgh \rb) - \lb( \bp \cdot \bk \rb) \lb( \bk \cdot \bgh\rb) \rb) $. Evaluating this integral with the Fourier space solution \eqref{eqn:density-Fourier-lim1}, \eqref{eqn:velocity-Fourier-lim1} gives
\begin{equation}
    \eta_{\rm{far}} = - \frac{Ri_v^{7/4} Pe^{-1/4}}{60 \lb(\beta^2 +2 \rb)} \lb( 2 E_E\lb( 1/\sqrt{2} \rb) - E_K \lb( 1/\sqrt{2} \rb) \rb) + O\lb( \lb( Ri_v Pe \rb)^{3/4}, Ri_v^{5/4} Pe^{1/4} \rb),
\end{equation}
where $E_K$ and $E_E$ are the complete elliptic integrals of the first and second kind. The potential energy changes and mixing in the far-field are caused by the concentration-dipole character of the swimming organism, which is why they are independent of the swimmer orientation and the propulsion type (pushers vs pullers). The error incurred in including the particle volume in the integral volume is $ O\lb( Ri_v Pe \rb) \ll \eta_{\rm{far}}.$

Accounting for both the near field and far field, the mixing efficiency is to leading order
\begin{equation}
    \eta = \eta_{\rm{near}} + \eta_{\rm{far}} = \frac{Ri_v \beta}{30 \lb( \beta^2 + 2\rb)} \lb( 3 \lb( \bp \cdot \bgh \rb)^2 - 1 \rb) - \frac{Ri_v^{7/4} Pe^{-1/4}}{60 \lb(\beta^2 +2 \rb)} \lb( 2 E_E\lb( 1/\sqrt{2} \rb) - E_K \lb( 1/\sqrt{2} \rb) \rb),
\end{equation}
where $\lb( 2 E_E\lb( 1/\sqrt{2} \rb) - E_K \lb( 1/\sqrt{2} \rb) \rb) \approx 0.8472$. The near-field mixing dominates over the far field if $Pe \gg Ri_v^3$. This means the near-field mixing dominates for a negligible concentration-dipole relative to the stresslet $\lb( Pe \gg Ri_v \rb)$. Conversely the far-field mixing dominates if $Ri_v^3 \gg Pe$. Also the overall mixing efficiency could be positive (or negative) as the swimmer could displace heavier fluid upwards (or downwards). The far-field mixing efficiency is negative while the near field one depends on $\bp$ and $\beta$, being also negative for $\beta \lb(3 \lb(\bp \cdot \bgh \rb)^2 - 1 \rb) < 0$.

Of particular importance is the mixing induced by vertically or horizontally swimming organisms, as swimmers in vertical density gradients display taxis (termed densitaxis) by rotating to swim along one of the two aforesaid orientations \cite{Shaik2024}. Pullers for instance rotate to swim vertically upwards (or downwards), but pushers swim horizontally. Then the mixing induced by vertically swimming pullers and horizontally swimming pushers, respectively, simplify to
\begin{align}
\eta_{{\rm{vertical}}} =& \, \frac{Ri_v \beta}{15 \left( \beta^2 + 2 \right)} - \frac{Ri_v^{7/4} Pe^{-1/4}}{60 \lb(\beta^2 +2 \rb)} \lb( 2 E_E\lb( 1/\sqrt{2} \rb) - E_K \lb( 1/\sqrt{2} \rb) \rb) \thickspace {\rm{and}} \\
\eta_{{\rm{horizontal}}} =& \, - \frac{Ri_v \beta}{30 \left ( \beta^2 + 2 \right)} - \frac{Ri_v^{7/4} Pe^{-1/4}}{60 \lb(\beta^2 +2 \rb)} \lb( 2 E_E\lb( 1/\sqrt{2} \rb) - E_K \lb( 1/\sqrt{2} \rb) \rb).
\end{align}
In the limit of negligible concentration-dipole relative to the stresslet $\lb( Pe \gg Ri_v \rb)$, far-field mixing is negligible, and $\eta_{\rm{vertical}}$ agrees with that found earlier \cite{Shaik2021}, $\eta_{\rm{vertical}} = Ri_v \beta/15 \lb( \beta^2 + 2 \rb)$.

We note that the limit $Pe \ll Ri_v^7$, where the mixing efficiency appears to be large, is not physically realizable for the biological swimming organisms in oceans, lakes, and ponds. Nevertheless, we can show that the mixing efficiency remains small for $Pe \ll Ri_v^7$. In this limit, while the mixing in the near-field is still given by Eq.~\eqref{eqn:mixing-near}, that in the far-field requires reexamination. By repeating the calculation in this subsection with $Pe = 0$, as applicable for $Pe \ll Ri_v^7$, we ultimately find that $\eta_{\rm{far}} \sim O\lb( Ri_v \rb)^{3/2} \ll \eta_{\rm{near}}$. Hence the mixing in the limit $Pe \ll Ri_v^7$ occurs primarily in the near-field of the swimmer, with the mixing efficiency given by Eq.~\eqref{eqn:mixing-near}.

We can also find the mixing caused by a dilute suspension of non-interacting squirmers that are homogeneously distributed in space but have random orientation. We find this mixing by taking an ensemble average of mixing due to a single squirmer
\begin{equation}
    \lb< \eta \rb> = \int{ \eta \Psi \lb( \bp \rb) } d\bp,    
\end{equation}
where $\Psi \lb( \bp \rb)$ gives the probability of finding a squirmer at any position and with orientation between $\bp$ and $\bp + d\bp$. For instance, isotropically oriented squirmers, $\Psi \lb( \bp \rb) = \frac{1}{4\pi}$, can only mix fluid through their far-field contribution, $\lb< \eta \rb> = \lb< \eta_{\rm{far}} \rb> $. The near-field mixing efficiency, being proportional to the second-degree Legendre polynomial $P_2 \lb( \bp \cdot \bgh \rb)$, vanishes upon integration due to symmetry. Without any near-field terms, we need to consider all the leading order terms in the far-field mixing efficiency, i.e., 
\begin{multline}
    \eta_{\rm{far}} = - \frac{Ri_v^{7/4} Pe^{-1/4}}{60 \lb(\beta^2 +2 \rb)} \lb( 2 E_E\lb( 1/\sqrt{2} \rb) - E_K \lb( 1/\sqrt{2} \rb) \rb) - \frac{5 \beta Ri_v^{5/4} Pe^{1/4}}{308\lb( \beta^2 + 2 \rb)} E_K\lb( 1/\sqrt{2} \rb) \lb( 3\lb( \bp \cdot \bgh \rb)^2 - 1 \rb) \\
    + \frac{3 \beta^2 \lb( Ri_v Pe \rb)^{3/4}}{2080\lb( \beta^2 + 2 \rb)} \lb( 2 E_E\lb( 1/\sqrt{2} \rb) - E_K \lb( 1/\sqrt{2} \rb) \rb) \lb( 21 + 24 \lb( \bp \cdot \bgh \rb)^2 + 11 \lb( \bp \cdot \bgh \rb)^4 \rb),
\end{multline}
as all these terms are of equal importance at $Ri_v/Pe \sim O\lb( 1 \rb)$.  Integrating $\eta_{\rm{far}}$ we derive the mixing by a suspension as
\begin{equation}
    \lb< \eta \rb> = - \frac{Ri_v^{7/4} Pe^{-1/4}}{60 \lb(\beta^2 +2 \rb)} \lb( 2 E_E\lb( 1/\sqrt{2} \rb) - E_K \lb( 1/\sqrt{2} \rb) \rb) + \frac{9 \beta^2 \lb( Ri_v Pe \rb)^{3/4}}{200 \lb( \beta^2 + 2 \rb)} \lb( 2 E_E\lb( 1/\sqrt{2} \rb) - E_K \lb( 1/\sqrt{2} \rb) \rb).
\end{equation}
Here the usual $O\lb(Ri_v^{7/4} Pe^{-1/4} \rb)$ part comes from the concentration-dipole character of swimmers, while the additional $O\lb( Ri_v Pe \rb)^{3/4}$ part comes from the force-dipole character of swimmers in stratified fluids and this contribution was also derived in the past by considering a dilute suspension of force-dipoles \cite{Wagner2014}.

But the aforesaid isotropic state evolves to an ordered state due to densitaxis. In this ordered state, pullers swim vertically upwards or downwards with equal probability $\Psi \lb( \bp \rb) = \delta \lb( \bp \cdot \lb( \bI - \bgh \bgh \rb) \rb)$ while pushers swim horizontally $\Psi \lb( \bp \rb) = \delta \lb( \bp \cdot \bgh \rb)$. Mixing caused by such suspension of vertically swimming pullers or horizontally swimming pushers is the same as that due to a single squirmer swimming along its steady state orientation, $ \lb< \eta \rb>_{{\rm{pullers}}} = \eta_{{\rm{vertical}}}$, $\lb< \eta \rb>_{{\rm{pushers}}} = \eta_{{\rm{horizontal}}}$.

\subsection{$l_s \gg l_{\rho}$}
In this limit, density transport is governed by advection in the outer region $r > l_s$. Hence we find the functional form of density in the matching region, $r \sim l_s \gg l_{\rho}$, by balancing the advection of disturbed density by particle velocity with the advection of background density
\begin{equation}
    Pe \bU_N \cdot \nabla \rho' \sim Pe \bu' \cdot \bgh.
\end{equation}
Using $\bu' \sim 1/r^2$, we get that $\rho' \sim 1/r$. We use this scaling to determine the stratification screening length by balancing the buoyancy forces with viscous forces near $r \sim l_s$
\begin{equation}
    \nabla^2 \bu' \sim Ri_v \rho' \bgh.
\end{equation}
We find that $l_s \sim Ri_v^{-1/3}$ and hence the condition $l_s \gg l_{\rho}$ implies that $Pe \gg Ri_v^{1/3}$.

We again rescale various quantities in outer region and solve the (rescaled) far-field equations in Fourier space to ultimately determine the far-field mixing efficiency. The rescaling proceeds by ensuring the rescaled variables are $O\lb( 1 \rb)$ in the matching region $\lb( r\sim l_s \rb)$. As the scaling of $l_s$ and $\rho'$ in this limit is different from that mentioned earlier in the other limit of $l_s / l_{\rho}$, rescaling here proceeds differently: $r = \rt/Ri_v^{1/3}$, $\rho' = Ri_v^{1/3} \rhot'$, $\bu' = Ri_v^{2/3} \but'$, and $p' = Ri_v \pt'$. Rewriting the far-field equations \eqref{eqn:FF-continuity}-\eqref{eqn:FF-density} in terms of rescaled variables, we derive
\begin{gather}
    \tilde{\nabla} \cdot {\tilde\bu}' = 0,
    \label{eqn:cont-rescaled2}\\
    - \tilde{\nabla} \tilde{p}' + \tilde{\nabla}^2 {\tilde\bu}' + \tilde{\rho}' \bgh + \bS \cdot \tilde{\nabla} \delta \left( \tilde\br \right) = \bzero,\\
    - \bU_N \cdot \tilde{\nabla} \tilde{\rho}' + {\tilde\bu}' \cdot \bgh = \frac{Ri_v^{1/3}}{Pe} \, \tilde{\nabla}^2 \tilde{\rho}' + \frac{Ri_v^{2/3}}{Pe} \, \bD \cdot \tilde{\nabla} \delta \left( \tilde\br \right).
    \label{eqn:density-rescaled2}
\end{gather}
Similarly, the far-field mixing efficiency expressed in terms of rescaled variables is
\begin{equation}
    \eta_{\rm{far}} = - \frac{Ri_v}{ 6 \pi \lb( \beta^2 + 2 \rb) } \int_{\tilde{\fV}}{ \rhot' \bgh \cdot \but' \tilde{dV}}.
    \label{eqn:mixing-far2}
\end{equation}

We neglect diffusion and concentration-dipole forcing for $Pe \gg Ri_v^{1/3}$ (or $l_s \gg l_{\rho}$) and solve the linear system of equations \eqref{eqn:cont-rescaled2}-\eqref{eqn:density-rescaled2} in Fourier space to derive the Fourier transform of density and velocity fields
\begin{align}
    \rhoh' \lb( \bk \rb) =& \, \frac{6\pi \beta \lb( \bp \cdot \bk \rb) \lb( k^2 \lb( \bp \cdot \bgh \rb) - \lb( \bp \cdot \bk \rb) \lb( \bk \cdot \bgh \rb) \rb) }{ k^4 \lb( \bp \cdot \bk \rb) + ik^2 - i \lb( \bk \cdot \bgh \rb)^2 }, \label{eqn:density-hat2}\\
    \bgh \cdot \buh' \lb( \bk \rb) =& \, \frac{ 6 \pi \beta i \lb( \bp \cdot \bk \rb)^2 \lb( k^2 \lb( \bp \cdot \bgh \rb) - \lb( \bp \cdot \bk \rb) \lb( \bk \cdot \bgh \rb) \rb) }{ k^4 \lb( \bp \cdot \bk \rb) + ik^2 - i \lb( \bk \cdot \bgh \rb)^2 }. \label{eqn:velocity-hat2}
\end{align}
We also extend the integration volume in \eqref{eqn:mixing-far2} to include particle volume and use convolution theorem to derive
\begin{equation}
    \eta_{\rm{far}} = - \frac{Ri_v}{48 \pi^4 \lb( \beta^2 + 2 \rb)} \int{ \rhoh' \lb( \bk \rb) \bgh \cdot \buh' \lb( - \bk \rb) d\bk }.
    \label{eqn:mixing-fourier2}
\end{equation}
Evaluating this integral with the Fourier space solution \eqref{eqn:density-hat2}, \eqref{eqn:velocity-hat2} yields that $\eta_{\rm{far}} = 0$. The error in including particle volume in the integration volume is $Ri_v \int_{\tilde{\fV_p}} { \rhot' \bgh \cdot \but' \tilde{dV} } \sim Ri_v \int_{\tilde{\fV_p}} { \rhot'_{\rm{match}} \bgh \cdot \but_{SS} \tilde{dV} } = 0$, meaning that the error $\ll \eta_{\rm{near}} \sim O\lb( Ri_v \rb)$. Here the density field in the matching region, $\rhot'_{\rm{match}} = \frac{3 \beta}{4 \rt^3} \lb( \lb( \bp \cdot \bgh \rb) \rt^2 + \lb( \brt \cdot \bgh \rb) \lb( \brt \cdot \bp \rb)  \rb)$, is found by taking the inverse Fourier transform of $\lim_{k \gg 1} \rhoh'\lb( \bk \rb)$.

Overall, the mixing efficiency to leading order in this limit is
\begin{equation}
    \eta = \eta_{\rm{near}} + \eta_{\rm{far}} = \eta_{\rm{near}} = \frac{Ri_v \beta}{ 30 \lb( \beta^2 + 2 \rb) } \lb( 3 \lb( \bp \cdot \bgh \rb)^2 - 1 \rb).
\end{equation}
Hence a vertically swimming puller or a horizontally swimming pusher, respectively, induce a mixing of
\begin{equation}
    \eta_{\rm{vertical}} = \frac{Ri_v \beta}{ 15 \lb( \beta^2 + 2 \rb) } \thickspace {\rm{and}} \thickspace \eta_{\rm{horizontal}} = \frac{Ri_v \beta}{ 30 \lb( \beta^2 + 2 \rb) }.
\end{equation}
Also a dilute homogeneous suspension of swimmers, with isotropic orientation distribution, cannot mix fluid at the leading order in $Ri_v$ and $Pe$, $\lb< \eta \rb> = 0$. But a suspension of vertically swimming pullers or horizontally swimming pushers can induce mixing identical to that caused by a single swimmer swimming along its steady state orientation, $\lb< \eta \rb>_{\rm{pullers}} = \eta_{\rm{vertical}}$ and $\lb< \eta \rb>_{\rm{pushers}} = \eta_{\rm{horizontal}}$.

\section{\label{sec:conclusions}Conclusions}
We found the mixing efficiency of a spherical squirmer swimming in vertical density gradients by assuming negligible inertia and weak density advection and stratification. The mixing efficiency was $O\lb( Ri_v \rb)$ in the near field of squirmer, while in the far-field, it was zero for $Pe \gg Ri_v^{1/3}$ and $O \lb( Ri_v^{7/4} Pe^{-1/4} \rb)$ for $Pe \ll Ri_v^{1/3}$. Here the P\'eclet number $Pe$ is the ratio of density advection to diffusion and the viscous Richardson number $Ri_v$ is the ratio of buoyancy to viscous forces. Although this mixing by a squirmer is finite and much larger than that induced by a force-dipole \cite{Wagner2014}, it is still small for weak stratifications $Ri_v \ll 1$. Also the mixing by a dilute homogeneous suspension of non-interacting squirmers, whether without any orientational order or with order caused by densitaxis, could be at most of the same order of magnitude as the mixing by a single squirmer. This work elucidates the effect of finite swimmer size on mixing, revealing that small swimmers could not induce significant mixing and thus do not contribute towards ocean mixing. Also that neglecting swimmer size and modeling swimmers as force-dipoles severely underestimates the induced mixing. 

We can estimate the near- and far-field mixing contributions for a few swimming organisms. We focus on vertically swimming organisms to obtain an upper bound on the near-field mixing. We also consider only the temperature stratifications for simplicity. For \textit{E. coli}, $Ri_v \approx 10^{-15} - 10^{-5}$, $Pe \approx 7\times 10^{-5} - 7\times 10^{-4}$, and as the squirming ratio $\beta < 0$, and the maximum value of $\left| \beta \right| \sim O\left( 1 \right)$ \cite{Berke2008}, we get the upper bounds on near- and far-field mixing, respectively, as $- \eta_{\rm{near}} \le 10^{-7}$ and $- \eta_{\rm{far}} \le 10^{-10}$. Also for \textit{Chlamydomonas reinhardtii}, $Ri_v \approx 10^{-13}-10^{-5}$, $Pe \approx 7 \times 10^{-4} - 7 \times 10^{-3}$, and $\beta \approx 2$ \cite{Gong2023}, hence we get that $\eta_{\rm{near}} \le 10^{-7}$, and $-\eta_{\rm{far}} \le 10^{-11}$. For \textit{Paramecium}, $Ri_v \approx 10^{-11} - 10^{-4}$, $Pe \approx 0.7$, and as $\beta < 0$ and assuming $\left| \beta \right| \sim O\left( 1 \right)$, we get that $-\eta_{\rm{near}} \le 10^{-6}$ and $\eta_{\rm{far}} = 0$ to the leading order in $Ri_v$ and $Pe$. For \textit{Volvox carteri}, $Ri_v \approx 10^{-10} - 10^{-3}$, $Pe \approx 0.21$, and $\beta \approx 0$, hence we get that $\eta_{\rm{near}} \approx 0$ to the leading order in $Ri_v$ and $Pe$, while $-\eta_{\rm{far}} \le 10^{-8}$.

These estimates can be used to assess the contribution of small swimming organisms towards ocean mixing. It is known that an approximate 6 TW of energy is expended for metabolic activities by all bacteria in the deep aphotic seas \cite{Dewar2006}. Assuming all of the 6 TW is spent for swimming and using the maximum positive mixing efficiency $\eta \approx 10^{-7}$, we get that an approximate 0.6 MW is spent in changing the potential energy of the fluid, which is still insignificant.

Unlike small or dilute suspension of organisms considered here, large or dense suspension of organisms could induce substantial mixing. Assuming our asymptotic analysis holds beyond its range of validity, particularly at strong stratifications $Ri_v \sim O\lb( 1 \rb)$, we expect large organisms experiencing strong stratification but with negligible inertia to induce significant mixing. This regime of strong stratification and negligible inertia has been overlooked in the literature, and could be explored with numerical simulations. Also a dense suspension of swimmers could induce significant mixing, as they are known to generate flows across length scales much larger than the individual swimmer size through collective motion. Understanding how the collective motion, and the induced mixing are affected by density gradients presents an intriguing avenue for future investigation.

\bibliography{references}

\end{document}

%% file: definitions.tex
\def \lb{\left}
\def \rb{\right}

\def \bgammad{\dot{\boldsymbol{\gamma}}}

\def \nablat{\tilde{\nabla}}
\def \bnabla{\boldsymbol{\nabla}}

\def \bOmega{\mathbf{\Omega}}

\def \rhob{\bar{\rho}}
\def \rhoh{\hat{\rho}}
\def \rhot{\tilde{\rho}}

\def \bsigma{\boldsymbol{\sigma}}

\def \bzero{\mathbf{0}}

\def \fB{\mathcal{B}}

\def \bD{\mathbf{D}}

\def \bF{\mathbf{F}}

\def \bI{\mathbf{I}}

\def \bL{\mathbf{L}}

\def \bS{\mathbf{S}}

\def \bU{\mathbf{U}}

\def \fV{\mathcal{V}}

\def \bX{\mathbf{X}}

\def \fh{\hat{f}}
\def \ft{\tilde{f}}

\def \bg{\mathbf{g}}
\def \bgh{\hat{\bg}}

\def \bk{\mathbf{k}}

\def \bn{\mathbf{n}}

\def \pt{\tilde{p}}
\def \bp{\mathbf{p}}

\def \rt{\tilde{r}}

\def \br{\mathbf{r}}

\def \brt{\tilde{\br}}

\def \bu{\mathbf{u}}

\def \buh{\hat{\bu}}

\def \but{\tilde{\bu}}

\def \bx{\mathbf{x}}